\documentclass{pasj00}

\begin{document}
\SetRunningHead{S. Deguchi et al.}{SiO Maser Sources near the Galactic Center}
\Received{2001/05/17}
\Accepted{2001/10/24}

\title{Observations of SiO Maser Sources within a Few Parsec 
from the Galactic Center}

\author{Shuji \textsc{Deguchi}%
  }
\affil{Nobeyama Radio Observatory, National Astronomical Observatory,\\
Minamimaki, Minamisaku, Nagano 384-1305}
\email{deguchi@nro.nao.ac.jp}

\author{Takahiro  \textsc{Fujii}
}
\affil{Institute of Astronomy,
The University of Tokyo, Mitaka, Tokyo 181-0015
}
\email{fujii@mtk.ioa.s.u-tokyo.ac.jp}

\author{Makoto  \textsc{Miyoshi}}
\affil{VERA Office, National Astronomical Observatory, 
Mitaka, Tokyo 181-8588
}
\email{miyoshi@miz.nao.ac.jp}

\and
\author{Jun-ichi {\sc Nakashima}}
\affil{Department of Astronomical Science, The Graduate University for Advanced 
Studies,\\
Nobeyama Radio Observatory, 
Minamimaki, Minamisaku, Nagano 384-1305}
\email{junichi@nro.nao.ac.jp}


%


\maketitle

\begin{abstract}
Mapping and monitoring observations of SiO maser sources 
near the Galactic center were made with the Nobeyama 45-m telescope
at 43 GHz. Rectangular mapping an area of approximately 
$200'' \times 100''$ in a 30$''$ grid,
and triangular mapping in a 20$''$ grid toward the Galactic center,
resulted in 15 detections of SiO sources; the positions of the sources
were obtained with errors of 5--10$''$, except for a few weak sources. 
Three-year monitoring observations 
found that the component at $V_{\rm lsr}=-27$ km s$^{-1}$
of IRS 10 EE flared to about 1.5 Jy during 2000 March--May, which was
a factor of more than 5 brighter than its normal intensity. 
Using the radial velocities and positions of the SiO sources,
we identified 5 which are counterparts of  previously observed
OH 1612 MHz sources. The other 10 SiO sources have no OH counterparts, 
but two were previously detected with VLA, and four are located
close to the positions of large-amplitude variables 
observed at near-infrared wavelengths. A least-squares fit to a 
plot of velocities versus Galactic longitudes gives a rather high
speed for the rotation of the star cluster around the Galactic center.
The observed radial-velocity dispersion is roughly consistent with 
a value obtained before. It was found that all of the SiO sources 
with OH 1612 MHz counterparts have periods of light variation 
longer than 450 days, while SiO sources without 
OH masers often have periods shorter than 450 days. This fact 
suggests that lower-mass AGB stars are more often 
detected in SiO masers than in the OH 1612 MHz line.
\end{abstract}

\section{Introduction}
The  Galactic-center star cluster consists of mixed
stellar populations (\cite{kra95}; \cite{mor96}). It involves a number
of late-type stars which are potential candidates for 
OH/SiO maser emitters. Deep surveys in OH 1612 MHz, H$_{2}$O 22 GHz 
and SiO 43 GHz masers (\cite{ sjo98b}b; \cite{men97}; \cite{izu98}) 
have been made; a dozen sources have been detected 
within a few parsec of the Galactic center. 
The accurate H$_{2}$O/SiO maser positions of these sources
observed by the Very Large Array (VLA) provided a precise alignment between
the near-infrared and radio coordinate frames 
(\cite{men97}), enabling the position of Sgr A*  to be pinpointed
with an accuracy better than 0.$''$1 
on near-infrared images. Detections of accelerating motions of
stars near the Galactic center fixed the mass of the cental object 
at about $3 \times 10^{6} M_{\odot}$ (\cite{ghe00}).
Because radio interferometers have a potential of
measuring the proper motions of stars relative to Sgr A*
[e.g.,  Sjouwerman et al.  (1998a)] more accurately than 
the present optical/infrared telescopes (\cite{gen96}; \cite{eck97}), 
to detect more SiO maser sources near the Galactic center and
to investigate their properties will be quite important. 

Using the  45-m telescope at Nobeyama, 
\citet{izu98} detected 14 SiO sources around the Galactic center.
These observations (by a beam width of about 40$''$) 
proved that the SiO maser source density is peaked at the Galactic center.
Because this was a set of pointed observations toward the Galactic center 
and two one-beam offset positions, the SiO source positions were not
determined with an accuracy better than the beam size. 
To remedy the positional uncertainties to some degree,  we made 
new mapping observations of SiO masers near the Galactic center 
using the 45-m telescope in the year 2000.
The present mapping observations on a 30$''$ grid 
can be used to derive the source positions with an uncertainty
of about 5--10$''$, depending on the signal-to-noise ratio.
Also, because the intensities of SiO masers are
expected to vary strongly on a time scale of one year, 
we monitored the intensities of SiO masers toward the Galactic center 
during 1999--2001.
During the three-year monitoring observations, we found
an SiO maser flare in the $-27$ km s$^{-1}$ component of IRS 10 EE
in 2000 March--June. We present the details 
of these observations in this paper.

\section{Observations}
Simultaneous observations of the SiO $J=1$--0, $v=1$ and 2 transitions at 42.122 
and 42.821 GHz, respectively, were made with the 45-m radio telescope at Nobeyama 
on 1999 June 8--12, 2000 May 20--29, and 2001 February 9. A cooled SIS 
receiver (S40) with a bandwidth of about 0.4 GHz was used, and the system temperature 
(including atmospheric noise) was 200--250 K (SSB). 
The aperture efficiency of the telescope was 
0.60 at 43 GHz. The half-power beam width (HPBW) was about 38$''$ at 43 GHz. 
A factor of 2.9 Jy K$^{-1}$ was used to convert the antenna temperature to flux 
density. An acousto-optical spectrometer array of low resolution (AOS-W) was 
used. Each spectrometer had 250 MHz bandwidth and 2048 channels, giving a 
velocity coverage of about 1700 km s$^{-1}$ and a spectral resolution of 1.7 
km s$^{-1}$ (per two binned channels). Observations were made in a position-switching mode, 
and the off-position was chosen to be 10$'$ away from the
Galactic center in azimuth. The telescope pointing was checked using a strong SiO 
maser source, OH $2.6-0.4$. The calibration of the telescope antenna temperature 
was made by observing the intensities of the $^{29}$SiO (thermal), 
H52$\alpha$ (recombination), U42.767, and SiO $J=1$--0, $v=1$ (maser) 
lines toward Sgr B2 MD5 [e.g., \citet{shi97}]. The total on-source integration time was
approximately 2 hours per day.

In the 1999 June observations, only the direction toward Sgr A* was observed;
(R.A., Decl., epoch)=($17^{\rm h}42^{\rm m}29.314^{\rm s}, -28^{\circ}59'18.3''$, 1950) 
(\cite{rog94}).
At this time, we spent three days looking for extreme velocity components
 ($|V_{\rm lsr}|>350$ km s$^{-1}$); no extreme velocity component above 0.04 
 and 0.03 K for the $J=1$--0, $v=1$ and 2 transitions, respectively,
was detected in the frequency range between 42.73--43.35 GHz
(approximate velocity span ranging from $-2200$ to 2700 km s$^{-1}$ 
for the SiO $J=1$--0 $v=1$ transition). In fact, several 
unidentified lines were found in the spectra toward Sgr B2 MD5 
in this frequency range: U42.767, U43.018, U43.026, and U43.178.
These U-lines should appear at $V_{\rm lsr}=$378 km s$^{-1}$ 
for the $J=1$--0 $v=2$ transition,
and $V_{\rm lsr}=723$,  668, and $-390$ km s$^{-1}$ for the  $J=1$--0 $v=1$ transition,
if the circumnuclear molecular ring (cf., \cite{wri01}) contains 
a sufficient number of molecules responsible for these U-lines. 
The spectra toward the Galactic center
above $|V_{\rm lsr}|>$ 350 km s$^{-1}$ were carefully checked  for  
contamination by these lines, but we found no such feature 
at the corresponding velocities (except for weak features due to 
$^{29}$SiO, H52$\alpha$, and U42.767 at several
$10'$ offset positions from Sgr A*). 

\begin{figure}
  \begin{center}
      \hspace{2cm}
    \FigureFile(75mm,180mm){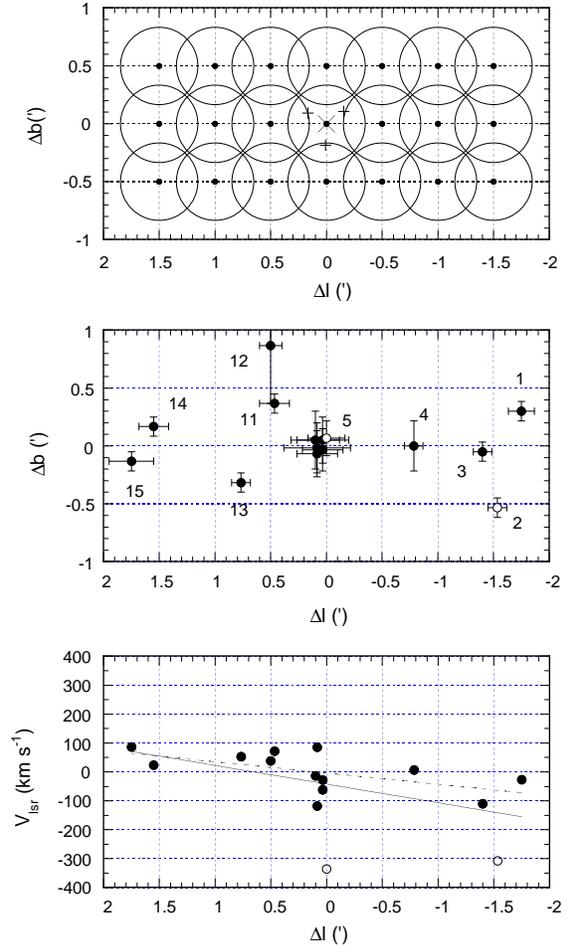}
  \end{center}
  \caption{Mapped positions of the telescope (top),  positions of the 15
  detected sources with error bars (middle), and position--velocity diagram of
  the detected sources (bottom). The coordinates, $\Delta l $  and $\Delta b$,
  indicate position differences from Sgr A* in Galactic coordinates in unit of 
  arcminute. The extreme sources with $|V_{\rm lsr}|>300$ km s$^{-1}$
  are shown as open circles in the middle and bottom panels. The numbers in the 
  middle panel correspond to the SiO source number in tables 1 and 2. 
  The solid and broken lines
  in the bottom panel are least-squares linear fits for all sources and for 
  all but the extreme sources, respectively.}\label{fig:l-b.map}
\end{figure}

In the 2000 May observations, we made mapping observations
in an area covering approximately $200''\times 100''$ centered toward Sgr A*.
Two different modes of mapping were used: 3-point (triangular) mapping
with 20$''$ separation toward Sgr A*, and 9-point ($3\times 3$ square) 
mappings toward $(\Delta l, \Delta b)=(0$ or $\pm 60''$, 0$''$) on a 30$''$ grid.
The 9-point mapping mode was utilized to obtain
a reasonably high level of signals in one run of the
typical mapping time of about two hours (approximately 10-minutes integration
per point).
The mapped points are shown in the top panel of figure 1.
The large circle in figure 1 indicates the effective beam of the telescope 
(HPBW $\sim 40''$). 
Because the telescope pointing has been known to be influenced
strongly by wind, we made 3-point mapping on those days with wind 
speed less than 5 m s$^{-1}$, when the telescope pointing was excellent
(error $\lesssim 5''$ accuracy), and  9-point mapping 
on relatively windy days with wind speeds of  5--10 m s$^{-1}$ 
(approximately $\lesssim  10''$ accuracy). Because the grid separation is
larger than HPBW in the 9-point mapping, the relative intensities of the SiO maser 
components in the grid points were supposed to be kept constant, 
even though the wind velocity was slightly high.
In the 2001 February observations, we could manage only a few hours of observation 
time, and we made observations in the 3-point mapping mode toward Sgr A*;
the signal-to-noise ratio of the spectra obtained 
was not very high. Therefore, we averaged the spectra of three 
positions (a half-beam width each away). They are 
shown at the bottom in figure 2.

A shallow survey of SiO sources 
in the Galactic center area was also made with the 45-m telescope 
using a multibeam receiver (S40M) prior to the present observations
(2000 March--April). 
This shallow observation detected 9 sources in the  $7'\times 13'$ area 
toward the Galactic center (\cite{miy01}). The present observations concentrated 
on a smaller and much closer area to the Galactic 
center with a longer integration time per point and
using a more sensitive single-beam receiver, S40.

As noted in \citet{izu98}, the  AOS-W spectra toward the strong continuum source, 
Sgr A* ($\sim 11$ Jy at 43 GHz; \cite{sof86} ; \cite{bec96}), exhibited a baseline 
distortion of about 0.3 K at the maximum,
and ripples due to a standing wave in the telescope system.
The ripples in the velocity range of $\pm$ 350 km s$^{-1}$ were relatively weak.
In order to remove these complex ripple features from the spectra, we took 
running means of the spectra (average of about 100 channels, or about 80 km 
s$^{-1}$ width), and the averaged spectra were subtracted from the originals. With 
this procedure, the baselines of the resulting spectra became quite flat. Because
the SiO maser lines are quite narrow (width less than 10 km s$^{-1}$) and weak 
($T_{\rm a} <0.2$ K), this method seems to work well, except for a strong line of $T_{\rm a}
\sim 0.5$ K at $-27$ km s$^{-1}$  (No. 6 in table 1). An additional baseline 
adjustment was made by taking a parabolic fit to the trough
between $\pm$ 40 km s$^{-1}$ near this strong feature.
The resulting three-year spectra toward Sgr A* [$(\Delta l, \Delta b)$=(0,0)] are shown 
in figure 2. 

Detections were judged by criteria similar to those given in \citet{izu98}.
All of the peaks for a single channel with $S/N>3$ and features for
several channels with  $(S/N)_{\rm broad} >5$ were treated as detection candidates. 
Then, all of the features were checked as to whether or not they were detected 
in the spectra of both the $v=1$ and 2 transitions of SiO,  
at nearby positions, or at different epochs of observations. 
Dubious features were discarded. 
After these careful checks, we selected 15 SiO spectral components 
as confident detections, which are listed in table 1; the number, 
telescope positions (relative to Sgr A*), $V_{\rm lsr}$, peak antenna temperatures,
integrated intensities, r.m.s. noise values, and signal-to-noise ratios
integrated over the emission profile, are given. We show the spectra of
the detected sources in  figures 3, 4, and 5. The components at $-117$ and
$-337$ km s$^{-1}$ were quite weak and only slightly above the critical level
of detections in 2000 May.
These components were detected at several different positions 
and epochs. They were, in fact, more clearly detected previously: 
the $-117$ km s$^{-1}$ component
in 1999, and the $-338$ km s$^{-1}$ component in 1997 (\cite{izu98}).
For the $-117$ km s$^{-1}$ component (No. 9) in figure 3,
the spectra taken in 1999 are shown.

\begin{figure}
  \begin{center}
      \hspace{2cm}
    \FigureFile(75mm,100mm){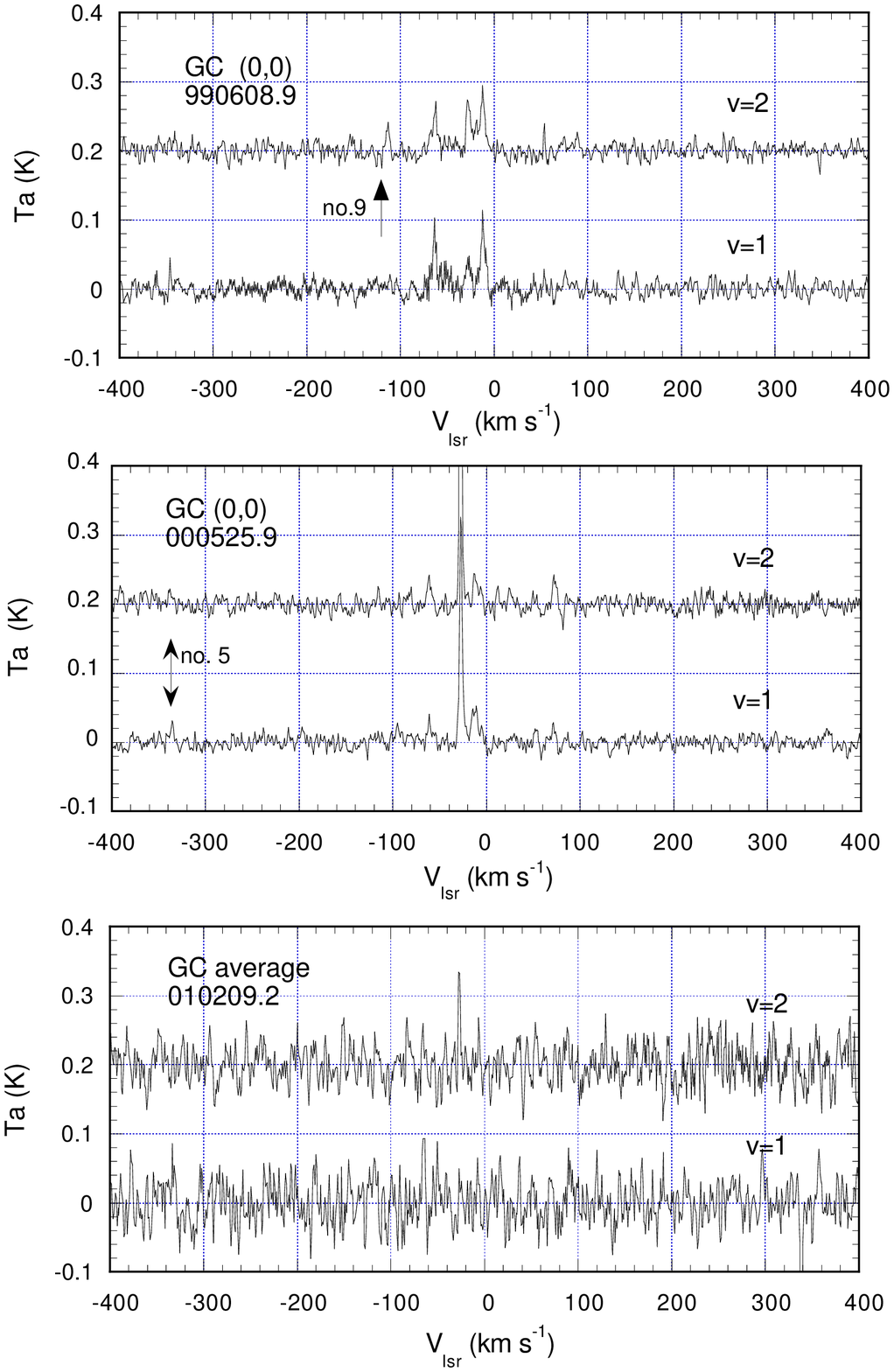}
  \end{center}
  \caption{Time variation of the SiO maser spectra toward Sgr A*. 
  The data were taken on 1999 June 8 (top), 2000 May 25 (middle), 
  and 2001 February 9 (bottom). In fact, the bottom spectra are averages 
 over the three 12$''$-offset spectra taken in the 3-point mapping mode to improve S/N.}
  \label{fig:time.variation}
\end{figure}
\begin{figure}
  \begin{center}
    \FigureFile(75mm,100mm){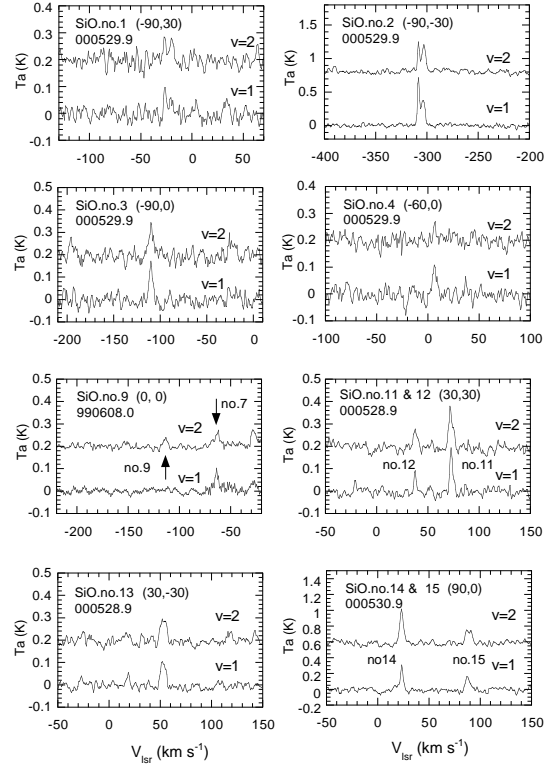}
  \end{center}
  \caption{SiO maser spectra of the detected sources. The source number shown upper left 
  of each panel corresponds to the number in tables 1 and 2. The observed positions 
  (in unit of arcsec from Sgr A*) are shown
  in parentheses.}\label{fig:spectra}
\end{figure}
\begin{table*}
  \caption{Observed Line Parameters.}\label{tab:table1}
  \begin{tabular}{rrrrrrrrrrrrl}
  \hline\hline
  & & \multicolumn{5}{c}{$J=1$--0, $v=1$}&\multicolumn{5}{c}{$J=1$--0, $v=2$}\\
 \cline{3-7} \cline{8-12} \\
No&($\Delta l, \Delta b)_{\rm obs}^{1}$&$V_{\rm lsr}$&$T_{\rm 
a}$&$F$&$T_{\rm rms}$&$S/N$
&$V_{\rm lsr}$&$T_{\rm a}$&$F$&$T_{\rm rms}$&$S/N$&Date\\
 &($''$, $''$)&{\scriptsize (km s$^{-1}$)}&(K)&{\tiny (K km s$^{-1}$)}&(K)& &{\scriptsize(km 
 s$^{-1}$})&(K)&{\tiny (K km 
 s$^{-1}$)}&(K)& &{\scriptsize (yymmdd)}\\
	\hline
1&($-$90,30)&$-$26.0&0.096&0.368&0.019&7.4&$-$27.4&0.084&0.496&0.021&7.2&000530\\
2&($-$90,$-$30)&$-$308.0&0.668&2.764&0.018&55.1&$-$308.3&0.403&2.173&0.021&33.3&000530\\
3&($-$90,0)&$-$110.2&0.172&0.700&0.019&13.6&$-$110.1&0.144&0.775&0.023&10.7&000530\\
4&($-$60,0)&6.3&0.109&0.436&0.018&8.8&7.4&0.087&0.296&0.014&8.5&000530\\
5&(0,0)&$-$335.8&0.031&0.123&0.009&5.2&$-$335.7&0.044&0.314&0.014&6.1&000525\\
6&(0,$-$12)&$-$27.8&0.349&1.228&0.008&59.8&$-$27.4&0.583&1.732&0.010&72.3&000526\\
7&(0,0)&$-$60.8&0.102&0.206&0.012&9.0&$-$62.0&0.109&0.243&0.014&8.6&000525\\
8&(0,-12)&85.2&0.073&0.265&0.008&12.4&84.8&0.046&0.103&0.010&5.0&000526\\
9&(10,6)&$\cdots$&$\cdots$&$\cdots$&0.009&$\cdots$&$-$117.3&0.036&0.160&0.010&5.5&000526\\
10&($-$10,6)&$-$12.5&0.051&0.157&0.010&6.9&$-$13.7&0.044&0.288&0.007&11.8&000526\\
11&(30,30)&71.7&0.302&1.027&0.022&18.8&71.3&0.249&1.151&0.028&14.4&000530\\
12&(30,30)&37.7&0.092&0.161&0.014&6.4&37.7&0.078&0.260&0.014&7.4&000530\\
13&(30,$-$30)&51.1&0.202&0.951&0.022&15.1&54.4&0.160&0.852&0.025&11.0&000531\\
14&(90,0)&23.3&0.311&1.037&0.021&20.5&23.2&0.416&1.690&0.026&24.3&000531\\
15&(90,0)&86.9&0.161&0.686&0.021&12.1&86.5&0.152&0.915&0.026&10.8&000531\\
\hline
\end{tabular}
$^{1}$ Observed telescope position.
\end{table*}

\begin{table*}
  \caption{Obtained positions and identifications to the previously detected objects.}\label{tab:table2}
  \begin{tabular}{rrrrrrrrllrr}
  \hline\hline
No.&$l$&$b$&$\Delta l$&$\Delta b$&$(\Delta l)_{\rm er}$&$(\Delta b)_{\rm er}$
&$V_{\rm lsr}^{\rm SiO}$&Ref$^{\ast}$&Identification$^{\dagger}$&$V_{\rm lsr}^{\rm Ref}$
&$\Delta r$\\
 & ($^{\circ}$) &($^{\circ}$) & ($''$)&($''$)&($''$)&($''$)& {\scriptsize (km s$^{-1}$)} &
 &{\scriptsize (km s$^{-1}$)}&{\scriptsize (km s$^{-1}$)}&($''$) \\
	\hline
1&359.915&$-$0.041&$-$105&18&7&5&$-$26.7&3&3--57&&6\\
2&359.919&$-$0.055&$-$92&$-$32&5&5&$-$308.1&1,3&359.918$-$0.055,3--2855&$-$307.9&5\\
3&359.921&$-$0.047&$-$84&$-$3&5&5&$-$110.1&2,3&M4&$-$105.1&\\
4&359.930&$-$0.045&$-$47&0&5&7&6.9&2,3&M3,3--88&17.5&6\\
5&359.944&$-$0.045&0&4&10&9&$-$335.8&2&C7&$-$341.9&\\
6&359.945&$-$0.047&2&-2&11&11&$-$27.6&1,2,4&359.946$-$0.048,C4&$-$26.4&6\\
&&&&&&&&&,IRS10EE&&\\
7&359.945&$-$0.045&2&3&14&12&$-$61.4&2  &C5 &$-$69.6&\\
8&359.946&$-$0.046&5&$-$1&18&13&85.0& & & & \\
9&359.946&$-$0.047&5&$-$4&12&8&$-$117.3&2,4&C6,IRS7&$-$121&7\\ 
10&359.946&$-$0.045&6&3&13&15&$-$13.1&2,4&C2,IRS15NE&$-$14.6&5\\
11&359.952&$-$0.040&28&22&8&5&71.5&1,2&359.954$-$0.041,P1&70.6&10\\
12&359.953&$-$0.032&30&52&6&30&37.7&3&3--885&&8\\
13&359.957&$-$0.051&46&$-$19&5&5&52.8&1,2,3,5&359.956$-$0.050,P2,3--5&48.5&8\\
14&359.970&$-$0.043&93&10&8&5&23.3&3&3--6&&3\\
15&359.973&$-$0.048&105&$-$8&12&5&86.7&1,3&359.970$-$0.049,3--3&88.8&8\\
\hline
\\
\end{tabular}
$^{\ast}$ References: 1---Sjouwerman et al. (1998b), 2---\citet{izu98}, 3---\citet{gla01},
4---\citet{men97}, 5---\citet{lev95}.\\
$^{\dagger}$ The names, 3--57, etc. (indicating the survey field and the star number)
refer to the large-amplitude variables in \citet{gla01}.
The names of the sources in \citet{izu98} are designated as 
Mn, Cn, and Pn, where M, C, and P stand the telescope positions
at ($-40''$,0), ($0''$, $0''$), and ($+40''$,0), respectively,
relative to Sgr A*, and n the number given to the detected SiO source 
in \citet{izu98}.  \\
\end{table*}

\section{Discussion}
\subsection{Source Positions and Identifications}
Because most of the components were detected at more than one telescope position,
we computed the most likely positions of the SiO features from the relative intensities
at the several observed positions. We assumed that antenna temperature
of a SiO maser component varies with the angular separation from the telescope pointing center 
according to the Gaussian beam shape with HPBW$=40''$ (which is taken to be slightly larger 
than the nominal HPBW because of pointing fluctuation due to wind). 
The most likely position of a component was calculated by minimizing the sum, $\Sigma 
[(F_{\rm obs}-F_{\rm exp})/T_{\rm rms}]_{i}^{2}$,
where the sum was made over the detected and undetected 
($F_{\rm obs}=0$) positions, $i$; $F_{\rm obs}$, $F_{\rm exp}$, 
and $T_{\rm rms}$ are the observed integrated intensity of the component,
the expected integrated intensity of the component 
at the observed position assuming the Gaussian beam 
shape, and the rms noise temperature, respectively. 
The errors in $\Delta l$ and $\Delta b$ are also calculated from
the distance which gives twice the minimum value of the above sum.
The resulting most-likely positions obtained are given in table 2;
the component number, Galactic longitude and latitude, the relative positions 
from Sgr A*, errors of positions,  
SiO radial velocity (averaged in $J=1$--0, $v=$1 and 2 
transitions), the reference and name of the previously detected sources, 
the radial velocity in the literature, and the separation of the positions
from the identified source are given. The obtained positions are plotted
in the middle panel in figure 1 with uncertainty bars.

The identifications were made by using both the positions
and radial velocities. A list of the previously detected OH 1612 MHz sources 
near the Galactic center (\cite{sjo98b}b) was used for
the identification. The positions for these OH sources are known
to an accuracy of better than $\sim 1''$. Previous identifications of SiO sources 
with OH sources were also made by \citet{izu98}.
No. 6 in table 2 is the OH/IR source, 359.946$-$0.048 (IRS 10 EE; \cite{men97}),
which has a radial velocity of $-27$ km s$^{-1}$. The location of this 
source is well known from near-infrared, OH, 
and SiO maser observations (\cite{men97}; \cite{ sjo98b}b). 
The position obtained in the present paper
agrees well with the OH position within an accuracy of about 6$''$.
The high signal-to-noise ratio of this component confirms 
that this method for determing the positions of SiO masers works nicely. 
Source No. 10 with $V_{\rm lsr}=-13$ km s$^{-1}$ 
has been identified as IRS 15 NE (\cite{men97}) with the VLA; 
the VLA position agrees with the position obtained in the present paper 
within an error of 5$''$.
The positions of the other 4 weaker sources (No. 2, 11, 13, and 15)
also coincide well with the positions of the identified OH sources 
within accuracies of 3--10$''$, thus
verifying this method for weaker sources. 

The radial velocity of SiO source No. 4, $V_{\rm lsr}=7$ km s$^{-1}$, is 
close to the center velocity of the OH 1612 MHz double peaks of
OH 359.931$-$0.050, (\cite{ sjo98b}b) at $V_{\rm lsr}=17.5$ km s$^{-1}$ 
(the expansion velocity is about 14 km s$^{-1}$).
Because the SiO position obtained is separated 
from the OH position by 11$''$, 
it is probably a different source. 
The  position of SiO source No. 8 (85 km s$^{-1}$)
agrees well with that of OH 359.947$-$0.046, which is known to
an accuracy of 4$''$
(but note that there is a large uncertainty in the SiO position). 
However, the radial velocity of the SiO masers, 85 km s$^{-1}$, 
is slightly outside of the
OH double-peak velocities (89.4 and 105.3 km s$^{-1}$).
Therefore, we left these two SiO sources unassigned.

It is not surprising that more than half of the  SiO maser sources
have no OH 1612 MHz counterpart.
Previous studies of 
SiO masers near the Galactic center (\cite{lin91})
and in the inner bulge (e.g., Deguchi et al.
2000a,b) revealed that approximately 2/3 of the SiO sources  
had not been detected previously in spite of a very sensitive 
OH 1612 MHz survey (\cite{lin92b}b; \cite{sev97}) with the VLA and the ATCA. 

Large-amplitude variables (including long-period and semiregular variables, 
as well as supergiants) are potential candidates for SiO maser emitters. 
We  compared the positions of SiO sources 
without OH identification with the known 
positions of the large-amplitude variables within 
12$'$ of the Galactic center (\cite{gla01}). 
The positions of these variables were measured
with the $K$-band array camera,  and are of a few arcsec accuracy.
We found that 4 SiO sources (No. 1, 4, 12, and 14) are located 
near to these long-period variables within the estimated positional uncertainty.
These identifications are given in columns 9 and 10 in table 2. 
Because the radial velocities 
of these large-amplitude variables are not known,
the identifications of these sources may be slightly 
less certain than the OH identifications.
So far, SiO sources No. 3, 5, 7, and 8 have no OH 1612 MHz counterpart
and no corresponding large-amplitude variable; nevertheless, these were 
detected before in SiO by \citet{izu98}.

We also checked the corresponding near-infrared objects 
in the 2MASS image server for unidentified sources. However,  
because the star density within 30$''$ of the Galactic center is too 
high [e.g., \citet{blu96}], it is quite difficult to identify 
the SiO sources in this region.
On the 2MASS images,  we could only find
one red candidate ($\sim$10 mag in the $K$-band) for source No. 3
within a 5$''$ error circle.
This red star, however, accompanies a faint extended ($\sim 5''$) feature
which is considerably elongated in galactic longitude, 
probably because of the coalescence of several stars due 
to the low spatial resolution of the 2MASS images.
\begin{figure}
  \begin{center}
    \FigureFile(75mm,100mm){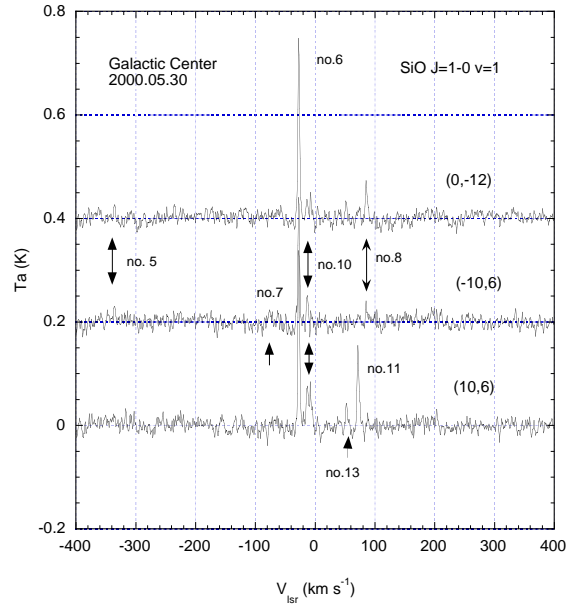}
  \end{center}
  \caption{SiO spectra in the $J=1$--0  $v=1$ transition
  at the three $12''$-offset positions around Sgr A*. The position offsets 
  from Sgr A* in units of arcsec are shown on the right side. The component number 
  in this figure corresponds to the number given in tables 1 and 2. 
  }\label{fig:three.V1}
\end{figure}
\begin{figure}
  \begin{center}
    \FigureFile(75mm,100mm){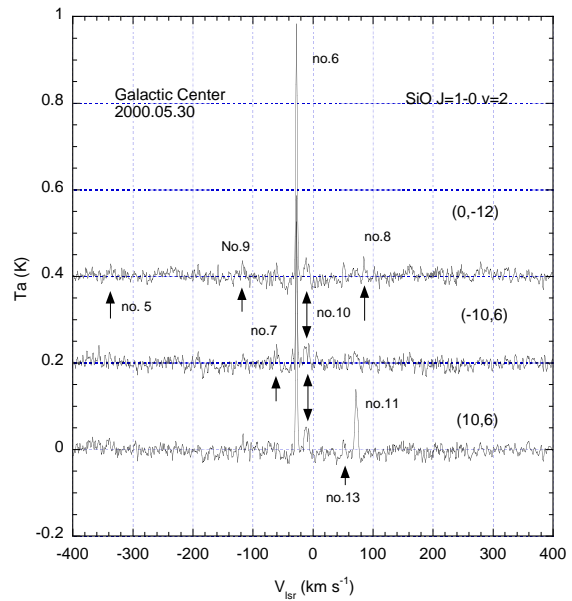}
  \end{center}
  \caption{Same as figure 4, but for the $J=1$--0, v=2 transition.
  }\label{fig:three.V2}
\end{figure}
\subsection{Velocity Distribution of the SiO Masers Sources}
The bottom panel of  figure 1 shows a longitude-velocity diagram
for the 15 detected sources. The positions given 
in table 2 were used for making this diagram.
A linear fit to the SiO radial velocities was made,
and the result is given in the first row of table 3. 
There are two extreme sources with
very large negative velocities, $-336$ and $-308$ km s$^{-1}$.
It is possible that the sources with very large radial velocities
are bulge stars with a very small angular momentum, 
rather than stars in the Galactic-center stellar cluster
(\cite{van92}; \cite{izu95}). They may be located far 
from the Galactic center in distance, but happen to be seen
near to it in projection, though this possibility is small.
We also made a linear fit excluding these two sources.
The results are also given in the second row of table 3.

The slope of about 1.1 ($\pm$0.5) km s$^{-1}$ per arcsec
(or 3890 km s$^{-1}$ deg$^{-1}$, or a rotation period 
of $2.3 \times 10^{5}$ yr around the Galactic center)  
is a factor of a few larger than the previously obtained values  
from SiO and OH maser observations on much larger scales 
[e.g., \citet{miy01}]. 
Even for the sample excluding 
the 2 extreme-velocity sources, the slope is
$\sim$2290 km s$^{-1}$ deg$^{-1}$.
The present value of the slope
is not compatible with the previously obtained low value 
by \citet{izu98} ($\sim$0.09 km s$^{-1}$ per arcsec), probably because
the positions of the sample in \citet{izu98} were not
known to an accuracy better than the telescope beam size 
of about $40''$; the positional accuracy  
was considerably better in our present observation.
The obtained high rotational speed of the SiO maser cluster
is, however, simply the slope of the least-squares fit, and 
must be interpreted very carefully.
Non-circular motions of stars and 
the gravitational potential made from 
the central compact mass and the nuclear star cluster
are responsible for the observed velocity structure.
The high rotational speed of these stars is comparable to
the rotational velocity of the circumnuclear gas ring 
which has been observed in various molecular lines 
[e.g., \citet{wri01}]. It is possible that these rapidly rotating 
AGB stars were born as a result of
star formation in the circumnuclear ring
(\cite{lev95}).

The velocity dispersion from the average linear fit
is approximately 108 km s$^{-1}$  
(or 54 km s$^{-1}$ excluding the extreme sources).
This value gives the mass
of the Galactic center (integrated to 100$''$; $\sim$ 4.1 pc) as
$\sim 1 \times  10^{7} M_{\odot}$,
when we use the virial theorem, and
is consistent with previous estimates of the mass 
of the Galactic center region
(\cite{kra95}; \cite{mor96}).
The radial velocity dispersion obtained for SiO maser sources, 110 km s$^{-1}$,
is slightly smaller than the value of 154 km s$^{-1}$ 
for the late-type stars 
within 12$''$ from the Galactic center (\cite{kra95}). 
In fact,  if we take the SiO sources only within a 20$''$ radius
from the Galactic center in our sample 
(No. 5--10; including the extreme source No. 5), 
we get a velocity dispersion of 133 km s$^{-1}$, 
giving a reasonable agreement with the result of \citet{kra95}. 

A simple judgement
on whether or not a particular star is dynamically bound to the
Galactic-center massive compact object can be obtained from
the characteristic binding energy per unit mass, 
$E_{\rm c}=(1/2)V_{\rm l.o.s.}^{2}-G M/r_{\rm p}$.
Here, $V_{\rm  l.o.s.}$, $r_{\rm p}$,
$G$, and $M$ are the observed radial velocity, the projected radius
from the galactic center,
the gravitational constant, and the mass of the central compact object
(for which we adopt $2.8\times 10^{6} \; M_{\odot}$; \cite{ghe00}; 
\cite{gen00}), respectively.
Considering a perpendicular velocity component and 
some depth along the line of sight, the characteristic binding energy, $E_{\rm c}$,
gives a lower limit to the real binding energy when the central mass dominates
the gravitational field.
For the SiO sources (No. 2, 3, and 15), the characteristic binding energies are
positive. Therefore, these sources are not dynamically bound to
the central compact object. 

Because the SiO intensity of the extreme object No. 5 was quite weak 
in the year 2000, the position uncertainty given in the present paper 
is quite large. The characteristic binding energy 
of object No. 5 could be positive if the true position is a few arcsec further
away from the Galactic center than that observed. The SiO intensity of this component   
was much stronger in 1997 (\cite{izu98}). Therefore, a more accurate
position should be obtained at a future date.

\begin{table*}
  \caption{Fitting results.}\label{tab:table3}
  \begin{tabular}{lccr}
  \hline\hline
Sample & Number  &  Best fit  & r.m.s.  \\
   &   of sources   &    (km s$^{-1}$) &(km s$^{-1}$) \\
\hline
  &&&\\
All sources &  15  & $-42.0(\pm 30.0) + 1.07 (\pm 0.51) [\Delta l$($''$)] & 108.4 \\
(including extreme sources)&&&\\
Low-velocity sources only &  13  & $-3.7(\pm16.3) + 0.67 (\pm 0.28)[\Delta 
l$($''$)]  & 53.9\\
(excluding extreme sources)  &&&\\
\hline
\\
\end{tabular}
\end{table*}

\subsection{Time Variation}
The intensities of SiO maser lines  change significantly on a time scale of one year.
Though the number of SiO sources detected in this study was similar to
the number given previously (\cite{izu98}), the two lists are not 
completely the same; 9 objects in the SiO spectra in 2000 are
inferred to be the same as those found in 1997
because of velocity coincidences.
The identifications are given in columns 9 and 10 of table 2.
If we compare figure 1 in the present paper with the middle panel of figure 1 
of \citet{izu98}, we can recognize significant variations in SiO maser intensities
during the last 5 years.  

We noticed in 2000 March that the SiO masers of source No. 6
(the $-27$ km s$^{-1}$ component, known as IRS 10 EE; \cite{men97}) 
had flared up to more than $T_{\rm a}\simeq$0.5 K ($\sim$ 1.5 Jy).
The SiO maser intensities of source No. 6  are plotted against time in figure 6.
We took the 1996 and 1997 data from \citet{men97} and \citet{izu98}, respectively,
and the 2000 March--April  data from \citet{miy01}. The peak and integrated 
intensities of the $-27$ km s$^{-1}$ component of \citet{miy01} 
agree quite well with the  2000 May results in this work.
Therefore, the flare lasted for more than  two months,
from the end of 2000 March to the end of 2000 May.
Because we did not observe the Galactic center in 1998,
it is not known if there was any brightening in 1998.
Near-infrared monitoring observations of 
this source over an interval of 4 years (\cite{woo98}; assigned as LWHM65) 
gave a period of 736 days for the light variation.
Extrapolating the fit of the light curve given
in \citet{woo98}, we estimate that the infrared maxima
of this source came around mid November of 1998 and  
late November of 2000. However, because the observed light curve
 (\cite{woo98}) has a large ambiguity, the estimated 
time of the light maximum is quite uncertain. It may have
occured in advance by several months before 2000 November 
(in fact, 2000 March--June), if we assume that the light  maximum
occurs at the same time as the SiO maser intensity maximum. 

The intensities of the SiO lines from IRS 10 EE, $T_{\rm a}\simeq 0.5$ K 
(about 1.5 Jy), in 2000 May were comparable to the intensity of 
the SiO $J=1$--0 $v=1$ maser from Sgr B2 MD5 (\cite{shi97}; \cite{mor92}). 
If they are scaled to a distance of 500 pc ($\sim 400$ Jy), 
they are comparable with the intensities of the Orion SiO masers.
Therefore, the SiO maser flare of IRS 10 EE may indicate that
the mass-loss rate of this star is temporarily enhanced to rates
comparable to those from Orion IRc2 and Sgr B2 MD5.
If the flare of IRS 10 EE is repeated periodically, it is probably associated
with a pulsation activity of the central star.
If it is irregular,  it is possible to consider another mechanism,
for example, a wind--wind collision between nearby massive stars
(\cite{yus92}). In addition, a tidal effect due to a close encounter 
between nearby stars might not be negligible, because 
the density of stars in the central star cluster is 
quite high. 

\begin{figure}
  \begin{center}
    \FigureFile(75mm,50mm){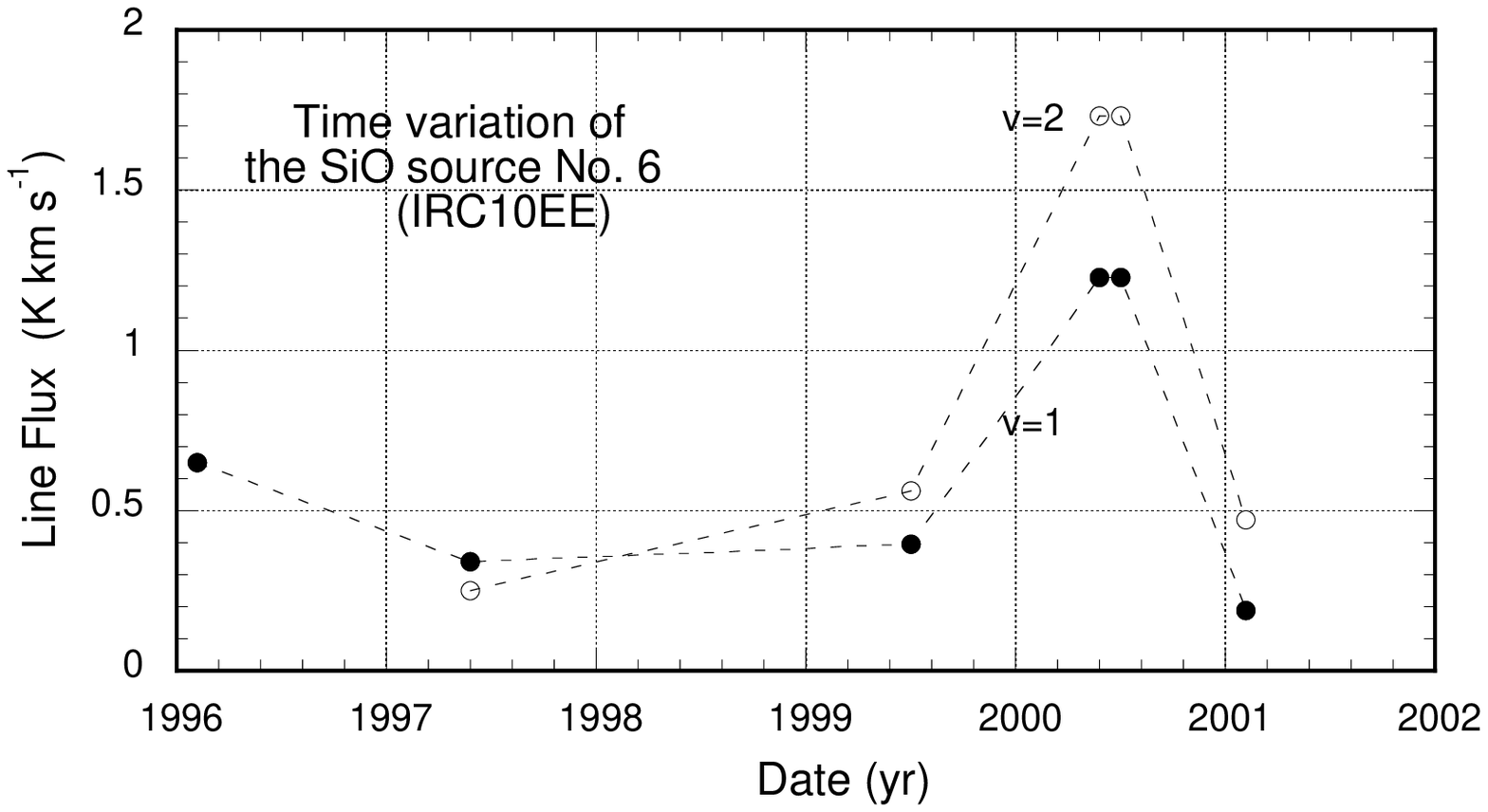}
  \end{center}
  \caption{Time variation of the SiO lines from No. 6 (IRS 10 EE).
  The data of 1996 and 1997 were taken from \citet{men97} and 
  \citet{izu98}. The data of 2000 March were taken from
  \citet{miy01}. The filled and unfilled circles indicate the
  SiO $J=1$--0, $v=1$ and 2 lines, respectively.
  The line fluxes in 1997 and 1999 were scaled by a factor of
  1.4 because IRS 10 EE is located at an offset of  
  about 10$''$ from  Sgr A*.  
  }\label{time.variation.IRS10}
\end{figure}

For the 8 SiO sources (No. 1, 2, 4, 6, 12, 13, 14, and 15),
the periods of light variation are known (\cite{blo98}; \cite{woo98}; \cite{gla01}).
A histogram of the periods is shown in figure 7. Because the sample 
involves only 8 stars, we divided it into two bins,
i.e., above and below 450 days. It is worth mentioning that
all sources with OH counterparts have periods of more than 450 days
(see  figure 7). Note that the average period of the sample 
with 412 large-amplitude variables
near the Galactic center given by \citet{gla01} is 427 days. 
Therefore, the Galactic-center SiO source sample is probably weighted more 
toward those variables with longer periods than the average. 
SiO masers are, however, occasionally detected in stars 
with shorter periods than those in which OH masers were found. 
Considering the mass--luminosity--period 
relation [e.g., \citet{vas93}], we conclude that  
lower-mass AGB stars are more often detected in SiO maser observations
than  in OH 1612 MHz observations. Because the size of the sample
is slightly small, the conclusion may not be free from statistical
fluctuations.  However, the present conclusion is quite
consistent with the finding by \citet{gla01} 
that the OH 1612 MHz sources have longer periods
than the average period in their sample. 
According to an SiO maser study 
of the Galactic disk IRAS sources (\cite{nak01}),
stars with bluer colors in terms of the IRAS 25/12 $\mu$m intensity ratio
(relatively thin dust envelope) tend to be detected 
in SiO masers more often than in OH 1612 MHz masers. This fact also seems to
agree qualitatively with the above result 
for the Galactic-center AGB stars, though the IRAS 25/12 $\mu$m colors 
do not necessarily correlate well with
the periods of AGB stars (\cite{whi91}; \cite{nak00}).

\begin{figure}
  \begin{center}
    \FigureFile(75mm,100mm){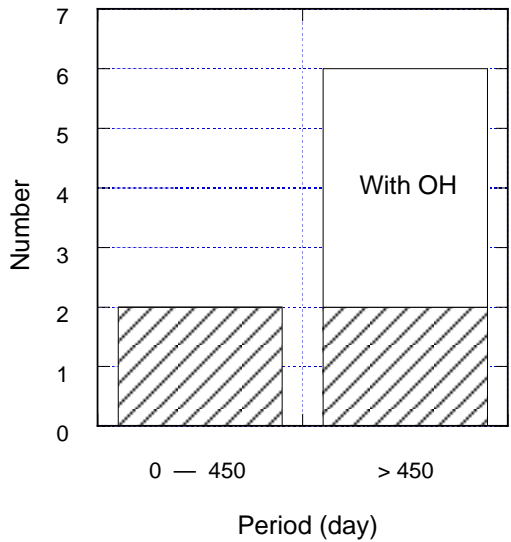}
  \end{center}
  \caption{Histogram of the periods for the SiO detected sources.
  The blank and shadow parts indicate the SiO sources with and without
  OH counterparts. All of the SiO sources with OH counterparts 
  are stars with periods longer than 450 days.
  }\label{P-histogram}
\end{figure}

The total number of SiO sources detected in the present paper is
15. If we include all of the SiO sources which were detected by \citet{izu98}
and in the present work, the total number of SiO maser stars 
in the region of 200$'' \times 100''$ from the Galactic center 
is 20.  Further monitoring observations of the SiO maser intensities of
the these Galactic center SiO sources are definitely required.

\section{Conclusion}
We have made mapping and monitoring observations of SiO maser sources 
near the Galactic center and have detected 15 SiO sources. Approximate
positions were obtained with accuracies of about 5--10$''$;
five sources were identified with the previously observed
OH 1612 MHz sources. Among the sources
without OH counterparts, four are close to the positions
of the large-amplitude variable stars observed at near-infrared
wavelengths and two to previously detected SiO sources
with accurate positions from the VLA. Three-year monitoring observations
of these objects found that SiO masers from IRS 10 EE flared up by a factor 
of more than 5 during March--May 2000. A least-squares linear fit of the
velocities to the Galactic longitude in the longitude--velocity diagram
gives a  high rotational speed for the star cluster 
around the Galactic center.

The authors thank I. Glass for stimulating discussions 
and reading the manuscript.
They also thank M. Morris, H. Izumiura, H. Imai, and A. Miyazaki for comments.  
This research was partly supported by Scientific Research Grant
(C2) 12640243 of Japan Society for Promotion of Sciences.



\end{document}